\documentclass[pra,10pt,superscriptaddress,footnoteinbib]{revtex4}
\usepackage{amsmath}
\usepackage{latexsym}
\usepackage{amssymb}
\usepackage{graphics,epstopdf}
\usepackage[colorlinks=true, citecolor=blue, urlcolor=blue ]{hyperref}
\usepackage{epsf,graphics,graphicx}
\usepackage{float}

\newcommand{\n}{\noindent}

\newcommand{\ed}{\end{document}}
\newcommand{\beq}{\begin{equation}}
\newcommand{\eeq}{\end{equation}}

\begin{document}
\title{Effect of non-uniform exchange field in ferromagnetic graphene}
\author{Debashree Chowdhury\footnote{Electronic address:{debashreephys@gmail.com}}${}^{}$  and B. Basu\footnote{Electronic
address:{sribbasu@gmail.com}} ${}^{}$} \affiliation{Physics and
Applied Mathematics Unit, Indian Statistical Institute, 203
B.T.Road, Kolkata 700108, India}


\begin{abstract}
\n
We have presented here the consequences of the non-uniform exchange field on the spin transport issues in spin chiral configuration of ferromagnetic graphene. Taking resort to the spin orbit coupling (SOC) term and non-uniform exchange coupling term we are successful to express the expression of Hall conductivity in terms of the exchange field and SOC parameters through the Kubo formula approach.
However, for a specific configuration of the exchange parameter we have evaluated the Berry curvature of the system. We also have paid attention to the study of SU(2) gauge theory of ferromagnetic graphene. The generation of anti damping spin orbit torque in spin chiral magnetic graphene is also briefly discussed. 
\end{abstract}

\maketitle
\section{Introduction}

Graphene \cite{gra} is a two dimensional (2D) layer of carbon atoms with unique flat geometry and remarkable intrinsic transport properties. Various unusual  properties of it make graphene an important material for the study of spin physics. The long spin flip length of graphene has made it a potential candidate for spintronic \cite{wolf,zutic} applications where the spin orbit coupling (SOC)  plays a very important role.   

In recent times, magnetized graphene has attracted a lot of attention from application standpoint. 
There are variety of ways for the experimental realization of magnetized graphene or more precisely graphene with spin imbalance. There may exist some intrinsic ferromagnetic correlations in graphene. Use of  an insulating ferromagnetic substrate or adding a magnetic material or magnetic dopants or defect on top of the graphene sheet may be other options to achieve ferro-magnetism in graphene. In particular, by depositing ferromagnetic insulating layer (FI) on graphene, magnetization is induced through the exchange proximity interaction (EPI). Induction of large exchange splitting has been demonstrated by depositing ferromagnetic insulator EuO on graphene \cite{123}. Thus induced magnetic effects in graphene basically rely on the principle of exchange interaction.  

Here we have considered the spin chiral nature of ferromagnetic graphene \cite{za}, where the sub bands of charge carriers with opposite spin direction are of electron or hole like. So there exists a regime
in which majority and minority spin carriers belong to different bands i.e conduction and valence bands. In this spin chiral configuration, the authors in \cite{za} demonstrate a different type of Hall effect from the spin
Hall mechanism in the extended Drude model approach.
The Hall effect that is generated by the spin Hall mechanism in graphene via the spin-orbit dynamics of the carriers can be well explored through non uniform exchange interaction \cite{1,sl,our}. As the gauge theory can be applied in describing transport in graphene \cite{a,b}, the role of momentum space and real space Berry curvature for a system with finite spin-orbit coupling is also worth studying.

On the other hand, in magnetic memory technologies there is a vital role of spin torque in magnetized switching via spin Hall effect. Spin transfer torque (STT), an important aspect of spintronics has been studied from the perspective of local magnetization dynamics. Recent theoretical and experimental studies suggest that the SOC can induce STT unlike the conventional spin injection  based STT. Simultaneous presence of SOC and local magnetization will induce a SU(2) gauge potential. Very recently, in ferromagnetic semiconductor an antidamping spin orbit torque (SOT) has been observed \cite{nature} which originates from the Berry curvature and provides an analogous understanding of intrinsic spin Hall effect via Berry curvature. Extension of this formalism for ferromagnetic graphene may provide some important consequences in the study of spintronic applications.         

In this paper, we would like to study the spin orbital dynamics of the carriers in spin chiral ferromagnetic graphene \cite{ex,za} within a framework of semiclassical theory. We find that the non-uniformity of the exchange coupling can be a potential candidate to explain various spin related aspects in spin chiral ferromagnetic graphene.

In case of spin Hall effect(SHE), the SOC provides the gateway to study different spin related issues. The well known Kubo transport formula has been used to study the Hall conductivity of spin orbit coupled system. We have explicitly used this approach to calculate the conductivity of a ferromagnetic graphene system, where the exchange coupling is momentum dependent. In our system the total conductivity arises from the SOC due to external electric field and non-uniform exchange field. It has been shown that the Hall conductivity can produce pure spin current in our magnetic graphene system.

Furthermore, the $\vec{k}$ dependence of the exchange coupling may induce a different kind of SOC. Due to the presence of these two kind of SOC, we have a SU(2) gauge field in the system, from which we can calculate the Berry curvature, which is different for the up and down electrons.
\vspace{.2 cm}

The paper is organized as follows: In sec II, the Hamiltonian for the ferromagnetic graphene is formulated. The derivation of Hall conductivity is shown in the Kubo formula approach in section III. In sec IV We have studied the spin gauge and curvature for a particular configuration of momentum dependence of the exchange field. Section V deals with the derivation of Berry curvature for coordinate dependent exchange coupling. The section VI includes the derivation of a anti-damping spin orbit torque, when the exchange field is considered as time dependent. We conclude in section VII with a discussion about the importance of the non-uniform exchange coupling.

\vspace{1 cm}

\section{The Hamiltonian}

We here consider the spin chiral configuration of the ferromagnetic graphene. In this spin chiral configuration the sub bands of charge carriers with opposite spin direction are of electron or hole like. When we consider such a configuration of ferromagnetic graphene with a substrate induced SOC, the corresponding Hamiltonian can be written as \cite{za}
\begin{equation}
H = v_{F}\vec{\alpha} . \vec{k} +E_F + \vec{\sigma}.\vec{h} + V(r) + \lambda_{G} [\vec{\sigma}\times\vec{\nabla}_{r}V(r)].\vec{k},
\end{equation}
where $\alpha$ is similar to Dirac matrix in spin space. $v_F$ is the Fermi velocity,  $\vec{\sigma}=(\sigma_x.\sigma_y,\sigma_z)$ is the Pauli matrix and the Fermi energy is given by $E_F$. The third term indicates the exchange Hamiltonian ($H_{ex} = \vec{\sigma}.\vec{h}$), which is present due to the interaction between the local magnetization of the ferromagnet and the surface Dirac fermions, where $\vec{h}$ is the exchange energy vector. The Hamiltonian in (1) is not in the form of a usual graphene Hamiltonian \cite{kane}. But here we have considered pristine spin chiral configuration of ferromagnetic graphene. There are other approaches \cite{jalil} of treating the ferromagnetic character in the Hamiltonian as well. But as we are interested in the spin chiral nature we can proceed with the Hamiltonian (1). The term
 $H_{ex}$ is similar to the Zeeman term which appears as a consequence of external magnetic field. In mean field theory one can consider this exchange field as an
effective magnetic field. The last term on the right hand side of the eqn. (1) is the SOC term, with $\lambda_{G}$ as the SOC strength. $V(\vec{r})$ represents the total potential in the system which includes external potential due to external electric field ($V_{e}$), potential due to static disorder ($V_{d}$) and crystal potential ($V_{0}$). In this context, the charge and spin Hall effect \cite{sh1,bc,cb,bcs} in spin chiral ferromagnetic graphene is discussed in \cite{za}. The exchange interaction thus induced can be non uniform \cite{1,sl} as well. 
The exchange vector may be a function of coordinate, momentum or time. The authors in \cite{1}, have experimentally demonstrated that the exchange coupling in case of the deposition of ferromagnetic insulators may be momentum dependent. On the other hand the spin lens is described in \cite{sl} with the help of space dependent exchange field. The non-uniform exchange coupling can produce interesting effects and can provide a window to investigate enormous important issues in the area of spin transport. 


\section{conductivity in ferromagnetic graphene}
The Hamiltonian of the system with  momentum dependent exchange field  can be written as  
\begin{equation} H = v_{F}\vec{\alpha} . \vec{k}+E_F + \vec{\sigma}.\vec{h}(\vec{k}) + V(r) + \lambda_{G} [\vec{\sigma}\times\vec{\nabla}_{r}V(r)].\vec{k}
\end{equation}
 Collecting only the dynamical terms, we can rewrite the Hamiltonian as
\begin{equation} H(\vec{k}) = E_G({\vec k})+ E_F + \vec{\sigma}. (\vec{h}+\vec{m})(\vec{k}) = \epsilon({\vec k}) + \vec{\sigma}.\vec{M}({\vec k}),\end{equation} where $\epsilon({\vec k})= E_G({\vec k})+E_F ,$
$\vec{m}(\vec{k}) = \lambda_G\vec{\nabla}_{r}V(r)\times \vec{k}$
 and $\vec{M}(\vec{k}) = (\vec{h}+ \vec{m})(\vec{k}). $  We do our analysis for exchange energy $\vec{\sigma}.\vec{h}(\vec{k})  <  $ Fermi energy $E_F,$  such that  $\alpha$ is equal to the unit matrix in the spin space and the carriers are electron like with up and down spin.
Here  $i = 1,2,3$ and  $\epsilon(k)$ is the kinetic dispersion energy. Also $\vec{m}(\vec{k}) = \lambda_G\vec{\nabla}_{r}V(r)\times \vec{k}$  
 and  $\vec{M}(\vec{k}) = (\vec{h}+ \vec{m})(\vec{k}) .$ 
 
In ferromagnet with a constant exchange energy a specific type of charge Hall effect is predicted \cite{za} that is generated by spin Hall mechanism in the absence of an external magnetic field. Within the semi-classical theory of spin-orbital dynamics of carriers, a longitudinal electric field produces a pure charge transverse current with no polarization of spin. 
 
Using the Kubo method \cite{kubo}, it is possible to derive the Hall conductivity from Hamiltonian (3). Here we consider the exchange vector as momentum dependent and our goal is to show that the conductivity is intimately connected to the ${\vec{k}}$ space Berry curvature, which is quite similar as the semiconducting system.   
 
Using the Matsubara Greens function technique, it is well known that it is possible to write the conductivity in the following form
\beq \sigma_{xy}=  \lim_{\omega \to 0}\frac{i}{\omega}Q_{xy}(\omega + i\delta),  \eeq
with \beq Q_{xy}(i\nu_{m}) = \frac{1}{C\beta}\sum_{k,n} tr\{J_{x}(\vec{k})G[k, i(\omega_{n} + \nu_{m})]J_{y}(\vec{k})G(\vec{k}, i\omega_{n})\} .\eeq  $G(\vec{k}, i\omega_{n})$ is the single particle Greens function, $\Omega$ is the system area, $J_{i}(\vec{k})$s are the current operator. $\omega_{n}$ and $\nu_{m}$ are the fermionic and bosonic Matsubara frequencies and can be expressed as $\omega_{n} =(2n +1)/\beta$ and $\nu_{m} = 2m\pi /\beta$. The single particle Greens function 
is given by 
\begin{eqnarray} 
G(\vec{k}, i\omega_{n}) &=& \left(i\omega_{n} - H(\vec{k})\right)^{-1}\nonumber\\ 
&=& \frac{1}{i\omega_{n} - \epsilon(k) - M_{i}(\vec{k})\sigma^{i}}\nonumber\\
&=& \frac{i\omega_{n} - \epsilon(k) + M_{i}(\vec{k})\sigma^{i}}{\left(i\omega_{n} - \epsilon(k) - M\right)\left(i\omega_{n} - \epsilon(k) + M\right)}\nonumber\\
&=&\frac{\frac{1}{2}[1 +\frac{M_{i}(\vec{k})\sigma^{i}}{M}]}{i\omega_{n}-\epsilon(\vec{k}) - M} + \frac{\frac{1}{2}[1 - \frac{M_{i}(\vec{k})\sigma^{i}}{M}]}{i\omega_{n}-\epsilon(\vec{k}) + M}\nonumber\\
&=& \frac{R_{+}}{i\omega_{n} - E_{+}(\vec{k})} + \frac{R_{-}}{i\omega_{n} - E_{-}(\vec{k})} 
\end{eqnarray}
where $M = \sqrt{M_{i}(\vec{k})M^{i}(\vec{k})},$ $R_{\pm} = \frac{1}{2}[1 + M_{i}(\vec{k})\sigma^{i}/M]$ and $E_{\pm}(\vec{k}) = \epsilon(\vec{k}) \pm M.$ Using the Hamiltonian the current operator $J_{i}(\vec{k})$ can be derived as  
\beq J_{i}(\vec{k}) = \frac{\partial H(\vec{k})}{\partial k_{i}} = \frac{\partial \epsilon(\vec{k})}{\partial k_{i}} + \frac{\partial M_{j}(\vec{k})}{\partial k_{i}}\sigma^{j},\eeq
where $i,j = x,y$ are the space indices. $Q_{xy}(i\nu_{m})$  can be defined as 
\beq Q_{xy}(i\nu_{m}) = \frac{1}{C\beta}\sum_{s,t=\pm}\sum_{\vec{k},n}tr\left[J_{x}(\vec{k})G(\vec{k},i(\omega_{n}+ \nu_{m}))J_{y}(\vec{k})G(\vec{k},i\omega_{n})\right],\eeq where $s$, $t$ are band indices and $n$ is the Matsubara index. We can now recast $Q_{xy}(i\nu_{m})$ in terms of the Fermi distribution function if we employ the Matsubara frequency sum over the indices $n$, and we can write $Q_{xy}(i\nu_{m})$ as \cite{fujita}
\beq Q_{xy}(i\nu_{m}) = \frac{1}{C}\sum_{s,t=\pm}\sum_{\vec{k}}\frac{tr[J_{x}(\vec{k})R_{s}(\vec{k})J_{y}(\vec{k})R_{t}(\vec{k})]}{i\nu_{m} -E_{s}(\vec{k}) + E_{t}(\vec{k})}(n_{t} - n_{s})(\vec{k}),\eeq
where $(n_{t} - n_{s})(\vec{k}) = (n_{F}(E_{t}(\vec{k})) - n_{F}(E_{s}(\vec{k})),$  and $n_{F}(\epsilon_{t}(\vec{k})) = \frac{1}{e^{\beta E_{t}} + 1}.$ Here $n_{F}(\epsilon_{s}(\vec{k})) = \frac{1}{e^{\beta (E_{s} - i\nu_{m})} + 1}$ denotes the Fermi distribution function for energy $E_{t}$ and $E_{s}$ respectively. 
The expression of $\sigma_{ij}$ can now be written as
\beq \sigma_{ij}=  \lim_{\omega \to 0}\frac{i}{\omega}Q_{ij}(\omega + i\delta)  .\eeq Inserting (9), in (10) we can write
\begin{widetext}
\begin{eqnarray}\label{si} 
\sigma_{xy} &=&  \lim_{\omega \to 0}\frac{i}{\omega}\frac{1}{C}\sum_{s,t=\pm}\sum_{\vec{k}}\frac{tr[J_{x}(\vec{k})R_{s}(\vec{k})J_{y}(\vec{k})R_{t}(\vec{k})]}{\omega + i\delta -E_{s}(\vec{k}) + E_{t}(\vec{k})}(n_{t} - n_{s})\nonumber\\
&=& -\frac{i}{C}\sum_{\vec{k}}\frac{tr[J_{x}(\vec{k})R_{-}(\vec{k})J_{y}(\vec{k})R_{+}(\vec{k}) - J_{x}(\vec{k})R_{+}(\vec{k})J_{y}(\vec{k})R_{-}(\vec{k}) ]}{(E_{+}(\vec{k}) - E_{-}(\vec{k}))^{2}}(n_{+} - n_{-})\nonumber\\
&=& -\frac{i}{C}\sum_{\vec{k}}\frac{tr[J_{x}(\vec{k})R_{-}(\vec{k})J_{y}(\vec{k})R_{+}(\vec{k}) -J_{x}(\vec{k})R_{+}(\vec{k})J_{y}(\vec{k})R_{-}(\vec{k}) ]}{4M^{2}}(n_{+} - n_{-})\nonumber\\
&=& -\frac{i}{C}\sum_{\vec{k}}\frac{D(\vec{k})}{4M^{2}}(n_{+} - n_{-}), 
\end{eqnarray}  
\end{widetext}

where $D(\vec{k}) = tr[J_{i}(\vec{k})R_{-}(\vec{k})J_{j}(\vec{k})R_{+}(\vec{k}) -J_{i}(\vec{k})R_{+}(\vec{k})J_{j}(\vec{k})R_{-}(\vec{k}) ].$ Explicitly in terms of the exchange vector we can write      
\begin{eqnarray}\label{ak}
 D(\vec{k}) 
 &=& \frac{1}{2}tr[(\frac{\partial \epsilon(\vec{k})}{\partial k_{x}} + \frac{\partial M_{\alpha}(\vec{k})}{\partial k_{x}}\sigma^{\alpha}) \frac{\frac{\partial M_{\beta}}{\partial k_{y}}M_{\gamma}}{M}(\sigma^{\beta}\sigma^{\gamma} - \sigma^{\gamma}\sigma^{\beta})]\nonumber\\
 &=& \frac{1}{2M}\frac{\partial M_{\alpha}(\vec{k})}{\partial k_{x}}\frac{\partial M_{\beta}(\vec{k})}{\partial k_{y}}M_{\gamma}tr\left(\sigma^{\alpha}\sigma^{\beta}\sigma^{\gamma} - \sigma^{\alpha}\sigma^{\gamma}\sigma^{\beta}\right)\nonumber\\
&=& \frac{2i\epsilon_{\alpha\beta\gamma}}{M}\frac{\partial M_{\alpha}}{\partial k_{x}} \frac{\partial M_{\beta}}{\partial k_{y}}M_{\gamma},
\end{eqnarray} 
where $\alpha, \beta,~\gamma$ are the space indices.
In this simplification, we  have used the relation  $tr(\sigma_{\alpha}\sigma^{\beta}\sigma^{\gamma} - \sigma_{\alpha}\sigma^{\gamma}\sigma^{\beta}) = 4i\epsilon_{\alpha\beta\gamma}.$
We should remind here that $\vec{M}(\vec{k}) = (\vec{h}+ \vec{m})(\vec{k}),$ which depends on the exchange vector as well as the SOC term. 
Thus the Hall conductivity in the x-y plane can be written as
\beq \sigma_{xy} = \frac{1}{2M^{3}C}\sum_{k}\epsilon_{\alpha\beta\gamma} \frac{\partial M_{\alpha}}{\partial k_{x}} \frac{\partial M_{\beta}}{\partial k_{y}}M_{\gamma}(n_{+} - n_{-})(\vec{k}).\label{34}\eeq 
The dependence of the Hall conductance on the exchange field $\vec{h}(\vec{k})$ is evident in (\ref{34}).     
Over the first Brillouin zone we can write the Hall conductivity in x-y plane as
\beq \sigma_{xy} = \frac{1}{2C}\int_{FBZ} \frac{c_{x}c_{y}}{4\pi^{2}}d^{2}\vec{k}\left(\vec{M}.\frac{\partial \vec{M}}{\partial k_{x}} \times\frac{\partial \vec{M}}{\partial k_{y}}\right),\eeq
where  $c_{x}c_{y} = C.$
From (14), we can definitely write the conductivity as 
\beq \sigma_{xy} = \frac{1}{8\pi^{2}}\int_{FBZ}d^{2}k \Omega_{z}(\vec{k}),\label{con}\eeq
where $\Omega_{z}(\vec{k})$ is the Berry curvature in momentum space.
One may note here that for $V(\vec{r})=V_0,$ where $V_0$ is the crystal potential (which is considered to be constant), the conductivity depends only on the inhomogeneity of the exchange vector in momentum space. The Hall conductivity is then given by
\beq
\sigma_{xy}= \frac{1}{2C}\int_{FBZ} \frac{c_{x}c_{y}}{4\pi^{2}}d^{2}\vec{k}\left(\vec{h}.\frac{\partial \vec{h}}{\partial k_{x}} \times\frac{\partial \vec{h}}{\partial k_{y}}\right),\eeq

In this context it is of interest that even in the absence of an electric field, momentum dependence of the exchange vector can induce Hall conductance in ferromagnetic graphene. In \cite{1}, the momentum dependence of exchange coupling is experimentally demonstrated. Our next goal is to choose a specific choice for the momentum dependence, which may produce some impressive results. 
  
The Berry curvature for spin Hall systems carries a spin dependence and is equal but opposite for spin up and down electrons i.e $\Omega_{z}^{\uparrow}(\vec{k}) = - \Omega_{z}^{\downarrow}(\vec{k})$. Thus eqn (\ref{con}) indicates that  
we have the spin Hall conductivity as \beq \sigma_{sH} = \sigma^{\uparrow}_{xy} - \sigma^{\downarrow}_{xy},\eeq whereas the charge conductivity is given by 
\beq \sigma^{c}_{xy} = \sigma^{\uparrow}_{xy} + \sigma^{\downarrow}_{xy}, \eeq
which is effectively zero. 
Thus for a ferromagnetic graphene system the non inhomogeneity in the exchange coupling can produce a pure transverse spin current with equal number of spin up and spin down electrons in our system. 

\section{Spin gauge and the curvature}

The choice of a specific form of the dependence of the exchange coupling $\vec{h}(\vec{k})$ on the momentum  may help us to elucidate the effective Berry curvature in the system from a field theoretical analysis.   
 A simple form of the momentum dependent coupling as \cite{sr} $\vec{h}(\vec{k}) = h~\hat{n}(k_{x}, -k_{y}, 0),$ when incorporated in the Hamiltonian (1) gives, 
\beq
H = v_{F}\vec{\alpha} . \vec{k} + E_{F} + h(\sigma_{x}k_{x} - \sigma_{y}k_{y}) + V(r) + \lambda_{G} [\vec{\sigma}\times\vec{\nabla}_{r}V(r)].\vec{k}.
\eeq
The second term reminds us about the well known Dresselhaus coupling in semiconductor. Although the mathematical structure of this coupling is similar to that of the Dresselhaus coupling, but the physical origin is far different from the original one. This new kind of SOC coupling originates from the momentum dependent exchange coupling of the ferromagnetic graphene, that has been experimentally demonstrated in \cite{1}.
The SOC coupling appears here is not because of the inversion asymmetry.
Let us now consider that the potential $V(\vec{r})$ is only due to an external electric field in the z direction, so that the relevant part of the above Hamiltonian can be written as 
\beq 
H = v_{F}\vec{\alpha} . \vec{k} + h(\sigma_{x}k_{x} - \sigma_{y}k_{y}) +  \alpha_{G}(\sigma_{y}k_{x} - \sigma_{x}k_{y}), \eeq 
where we take $v_{F} = \hbar = 1$.
A straight forward simplification gives,  
\beq H = \alpha.p + \left(hk_{x} - \alpha_{G}k_{y}\right)\sigma_{x} + \left(\alpha_{G}k_{x} - hk_{y} \right)\sigma_{y} = \alpha.(\vec{p}- \vec{A}_{G}),\eeq
where $\vec{A}^{G}_{i}$ is the spin dependent gauge field induced by the spin orbit coupling and exchange coupling terms and $\alpha_{G} = \lambda_{G}E_{z},$ with $E_{z} = -\nabla_{z}V(\vec{r})$ 

Explicitly the gauge fields acting in the $x$ and $y$ direction are given by 

\begin{eqnarray}
A_{x}^{G} = -i\left(-hk_{y} + \alpha_{G}k_{x}\right)\sigma_{x}\\
A_{y}^{G} = i\left(hk_{x} - \alpha_{G}k_{y}\right)\sigma_{y}
\end{eqnarray}
The SU(2) gauge field $\vec{A}^{G}_{i}$ is spin dependent and non-Abelian by nature. From the field theoretical consideration this gauge field can generate a physical field through the curvature term. The  physical field generated due to the presence of the gauge $\vec{A}^{G}$ is given by ($\hbar$ = 1)
 \beq \Omega^{G}_{z} = \Omega^{G}_{x ~y} = \partial_{x}A^{G}_{y} - \partial_{y}A^{G}_{x} -i\left[A^{G}_{x}, A^{G}_{y}\right] \eeq
It is quite obvious that the $z$ component of the $\vec{k}$ space curvature only exists and  is given by 
\begin{eqnarray}
\Omega^{G}_{z}(\vec{k}) &=& 2\left(\left(-hk_{y} + \alpha_{G}k_{x}\right)\left(hk_{x} - \alpha_{G}k_{y}\right)\sigma_{z}\right)\\
&=& \pm 2\left(\alpha_{G}h k^{2} - k_{x}k_{y}(h^{2} +\alpha_{G}^{2})\right),
\end{eqnarray}
where $ k^{2} = k_{x}^{2} + k_{y}^{2} $ and $\pm$ indicates the two spin species of the carriers. The expression clearly shows that the curvature is different for the up and down electrons. 
 The variation of this curvature with  $k_{x}$ and $k_{y}$ is plotted in figure (2) (a).   
\begin{figure}
\includegraphics[width=6.0 cm]{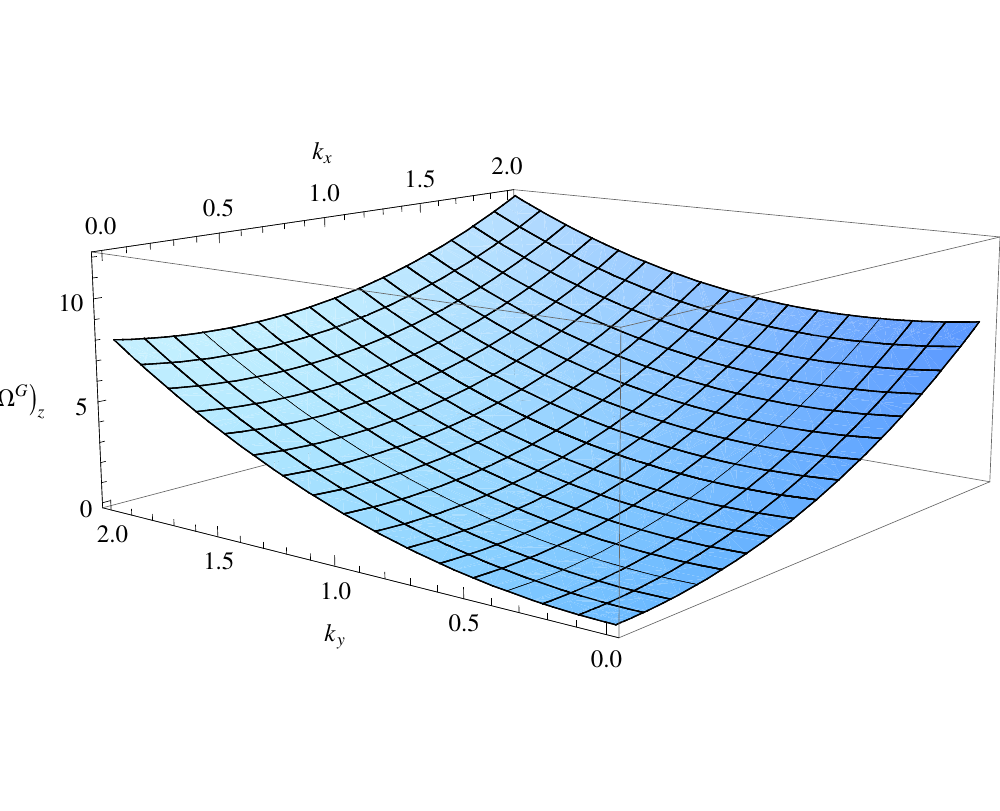}
\caption{ (Color online) Variation of curvature with momentum.}
\end{figure}
Using equations (15) and (26) we can calculate the conductivity in case of a momentum dependent exchange coupling as
\beq \sigma_{xy}= \frac{\pi\alpha_{G}h}{3a^{3}},\eeq
where $a$ is the lattice spacing. Thus the conductivity of the ferromagnetic graphene depends not only on the SOC coupling parameter but also on the exchange parameter as well.

\section{Real space Berry curvature for in-homogeneous exchange coupling}
In the previous section we have considered a specific configuration of the exchange vector when it is momentum dependent. Now our goal is to observe the consequences when the exchange field is coordinate dependent, which is already employed in the study of spin lens configuration \cite{sl, our}.
For an non-uniform exchange vector the Hamiltonian for ferromagnetic graphene (without any external parameters)can be written as 
\begin{equation}
H = v_{F}\vec{\alpha} . \vec{k} + \vec{\sigma}.\vec{h}(\vec{r})
\end{equation}
For the sake of simplicity the spin orbit coupling (SOC) term 
is also not considered. Actually, our motivation is to investigate the role of non-uniform exchange vector on real space berry curvature.
Variation of the exchange field can induce spin Berry gauge as follows
\begin{equation} {\cal{A}}_a^{\uparrow\downarrow} = -i\left\langle \uparrow\downarrow,\vec{h}(r)|\frac{\partial}{\partial r_a}|\uparrow\downarrow,\vec{h}(r)\right\rangle = \frac{\partial h_{a}(\vec{r})}{\partial r_a}A^{\uparrow\downarrow}_{a}(h),
\end{equation}
which can be rewritten as 
 \begin{equation}
 A^{\uparrow\downarrow}_{a}(h) = \left\langle \uparrow\downarrow,\vec{h}|\frac{\partial}{\partial h_{a}}|\uparrow\downarrow,\vec{h}\right\rangle
 \end{equation}
where $a=i,j,k $ are the space indices. $A^{\uparrow\downarrow}_{a}(h)$ is the exchange field dependent Berry gauge field appearing  due to the inhomogeneity of the exchange field vector. 

A physical field can be generated due to the presence of the gauge
$A^{\uparrow\downarrow}_{a}(h).$ 
We only have the $z$component of Berry curvature as
\begin{eqnarray}
\Omega_{c}(\vec{r}) = \frac{\partial {\cal{A}}^{\uparrow\downarrow}_{y}}{\partial x} - \frac{\partial {\cal{A}}^{\uparrow\downarrow}_{x}}{\partial y} = \frac{\partial h_{a}}{\partial x}\frac{\partial h_{b}}{\partial y}(\frac{\partial A^{\uparrow\downarrow}_{b}}{\partial h_{a}} - \frac{\partial A^{\uparrow\downarrow}_{a}}{\partial h_{b}})
\end{eqnarray}
For further analysis, in a standard notation of unit vector we write the exchange vector as 
\begin{equation}
 \vec{h} = h(sin\theta cos\phi, sin\theta sin\phi, cos\theta)
 \end{equation} 
 where  $\theta$ is the polar angle and $\phi$~is the azimuthal angle. 
 
 In the adiabatic approximation, the carriers remain in the same spin eignestates for $h(r_1)$and $h(r_2),$ and the flipping  between states is forbidden. Thus we can write the adiabatic gauge as
\begin{equation}A_{ad}^{\uparrow\downarrow}(h)=\pm\frac{1}{2}(1-\cos\theta)\nabla_h\phi
 \end{equation}
 where  $\pm$ denotes $\uparrow$ eigenstate parallel ($\downarrow$  anti- parallel) to $h(\vec{r})$. Thus the ${\vec r}$ space Berry curvature is given by, 
  \begin{equation}
\Omega_{c}(\vec{r})= \frac{\partial h_{a}}{\partial x}\frac{\partial h_{b}}{\partial y} \epsilon_{abc}(\pm \frac{h_{c}}{h^{3}}).
\end{equation}
In the adiabatic approximation, for the coordinate dependent exchange vector a physical field is generated 
which is the well known Berry curvature. This expression of Berry curvature shows that it is the source of monopole, if they exists.  

Now for a simplest dependence of the exchange coupling i.e $\vec{h} = (x, y , z)$ we can write 
the final form of Berry curvature as 
\beq \vec{\Omega} = \pm\frac{1}{2}\frac{\vec{h}}{h^{3}},\eeq
which is of nothing but the field generated due to magnetic monopole at the origin. But in general eqn. (31) can be written as 
\beq \Omega_{z}(\vec{r}) = \pm \frac{1}{2h^{3}}\vec{h}. (\frac{\partial \vec{h}}{\partial x}\times \frac{\partial \vec{h}}{\partial y} ).\eeq 

This is the expression of Berry curvature when we have the exchange field as space dependent. 
Following \cite{bruno} we can write the Drude Hall conductivity as
\beq \sigma_{xy} = \frac{ne^{3}\tau}{m^2}\Omega_{z}(\vec{r}),\eeq where n is the electron density and $\tau$ is the mean free time.
Thus with the help of non-uniformity of the exchange field we can successfully give a unified picture of conductivity for a SOC Hamiltonian.

\section{Anti-damping spin orbit torque} 
Apart from the above results we can argue that besides momentum and space dependence, the exchange field can be time dependent as well. In that case we can describe the antidamping spin orbit torque \cite{tf} in spin chiral configuration of ferromagnetic graphene. This can be an interesting issue regarding construction of magnetic memory based devices. To calculate the spin orbit torque we can start with equation (1) but with a time dependent exchange field. The torque should carry other contributions from divergence of spin current \cite{gd}. But here we are restricting ourself to the anti-damping torque only due to the exchange coupling term.
We can split the Hamiltonian into three parts as 
\beq H = H_{0} + H_{ex} + H_{1}, \eeq where $H_{0} = V(r) = e\vec{E} . \vec{r}, $ and $H_{ex} =  J\vec{\sigma}.\vec{h}(t) ,$ and $H_{1} = v_{F}\vec{\alpha} . \vec{k} - g\vec{\sigma} . \vec{B}_{SOI}(\vec{k}),$ where $\vec{B}_{SOI}(\vec{k})$ is the magnetic field appearing in the rest frame of the electron due to the spin orbit coupling and $J$ is the exchange coupling parameter. Let us consider the electric field in the x direction and the exchange field acts in the opposite to the external electric field i.e $-\vec{h}\parallel\vec{E}.$
The exchange field induces an additional magnetic field $B_{ex} = (2Jh, 0, 0)$ in the equilibrium. If we consider  $B_{ex}\gg B_{SOI},$ then we can write the component of spin in the out of plane z direction as \cite{nature} 
\beq s_{z} \approx \mp \frac{\alpha_{G} e E_{x}}{2J^{2}h^{2}} \eeq
From this equation it is clear that the out of plane spin vector depends not only on the SOC parameter but also on the exchange field as well. Also it depends on the direction of the exchange field with respect to the external electric field. It is maximum when $\vec{h}$ and $\vec{E}$ are anti parallel and zero for parallel orientation. In this generalized picture equation (39) can be written as 
\beq s_{z,h} \approx \mp \frac{\alpha_{G} e E_{x}}{2J^{2}h^{2}}cos\theta_{h,E}
.\eeq 
Integrating $s_{z,h}$ over all occupied states, the total non-equilibrium spin polarization can be obtained as
\beq S_{z} = 2gJhs_{z,h}
,\eeq where $g$ is the density of states. This non-equilibrium out of plane spin polarization produces a field in the out of plane direction which exerts a torque on the in plane exchange field. The expression of the torque can be written as 
\beq \frac{d\vec{h}}{dt} = \frac{J}{\hbar}(\vec{h}\times S_{z} \hat{z} ) \approx \vec{h}\times \left((\vec{E}\times \hat{z})\times \vec{h}\right). \eeq
The spin orbit torque can thus be obtained in terms of the exchange field. This can be studied in future. This aspect of magnetized graphene may play important role for using magnetized graphene as a future spintronic material. Furthermore, following \cite{nature} one can argue that the origin of this torque is actually the Berry curvature. The final expression of the out of plane spin polarization (non-equilibrium), which is basically responsible for the antidamping spin orbit torque can be written as 
\beq S_{z} = \frac{1}{V}\sum_{\vec{k}}(n_{+} - n_{-})\Omega_{z},\eeq
where $n_{\pm}$ is the Fermi distribution function, V is the volume and $\Omega_{z}$ is the z component of the Berry curvature. 
It is hoped that in future this preliminary  analysis on the spin orbit torque will be extensively studied and provide some impetus to work on the future spintronic memory based devices with ferromagnetic graphene. The beauty of this analysis is that it helps us to feel the importance of Berry curvature in analyzing the anti-damping spin orbit torque(SOT) in ferromagnetic graphene. One can simply argue that the origin of this SOT is nothing but the Berry curvature.  
 
\section{Discussion}
Induced magnetism in graphene is a gateway of enormous new physics and applications. By depositing EuO on top of the graphene one can control the exchange coupling through the molecular beam epitaxy (MBE) method. In this paper taking into account the non-uniform exchange field we have theoretically derived the expression of Berry curvature in the gauge theoretical approach. For a comprehensive review on gauge theory in spintronics see \cite{Fujita}. The non-uniformity might be due to coordinate dependence or momentum dependence. Both cases are studied here. With the help of Kubo formula and the Matsubara Greens function techniques we have analytically calculated the expression of Hall conductivity in case of a spin chiral ferromagnetic graphene. In the final form of the conductivity we clearly can visualize the dependence of the exchange field on Hall conductivity. Consideration of a particular form for the non-uniformity we can achieve a synthetic SOC field in the system, which has recent potential applications. Furthermore, the time dependence of the exchange vector helps us to explore a new arena of the physics behind the spin orbit torque in ferromagnetic graphene. The underlying origin of this torque is nothing but the the Berry curvature. In this context, it is of worth mentioning that there would be other components of the torque \cite{gd} due to the divergence of spin current. But here we are focusing on the torque due to the exchange coupling only. 

The experimental demonstration \cite{1} of the dependence of the exchange vector on the momentum inspired us to make the present analysis. Our gauge theoretical analysis shows the importance of the non uniform exchange vector in the study of various aspects of spin related issues in spin chiral ferromagnetic graphene . 

\end{document}